\documentclass[a4paper]{article}

\usepackage{INTERSPEECH2019}
\usepackage{hyperref}
\usepackage{enumitem}
\usepackage{graphicx}
\usepackage{multirow}
\usepackage{multicol}

\newcommand{\x}{\mathbf{x}}
\newcommand{\z}{\mathbf{z}}

\newcommand{\E}{\mathbb{E}}
\newcommand{\thetab}{\boldsymbol{\theta}}
\newcommand{\phib}{\boldsymbol{\phi}}

\makeatletter
\def\blfootnote{\xdef\@thefnmark{}\@footnotetext}
\makeatother

\title{Using generative modelling to produce varied intonation for speech synthesis}

\name{Zack Hodari, Oliver Watts, Simon King}
\address{The Centre for Speech Technology Research, University of Edinburgh, United Kingdom}
\email{\{zack.hodari, oliver.watts, Simon.King\}@ed.ac.uk}

\begin{document}

\maketitle

\begin{abstract}

    Unlike human speakers, typical text-to-speech (TTS) systems are unable to produce multiple distinct renditions of a given sentence. This has previously been addressed by adding explicit external control. In contrast, generative models are able to capture a distribution over multiple renditions and thus produce varied renditions using sampling.

    Typical neural TTS models learn the average of the data because they minimise mean squared error. In the context of prosody, taking the average produces flatter, more boring speech: an ``average prosody''. A generative model that can synthesise multiple prosodies will, by design, not model average prosody.

    We use variational autoencoders (VAEs) which explicitly place the most ``average'' data close to the mean of the Gaussian prior. We propose that by moving towards the tails of the prior distribution, the model will transition towards generating more idiosyncratic, varied renditions.

	Focusing here on intonation, we investigate the trade-off between naturalness and intonation variation and find that typical acoustic models can either be natural, or varied, but not both. However, sampling from the tails of the VAE prior produces much more varied intonation than the traditional approaches, whilst maintaining the same level of naturalness.

\end{abstract}

\section{Introduction}  \label{sec:introduction}

Prosody in natural human speech varies predictably based on contextual factors. However, it also varies arbitrarily, or due to unknown factors \cite{xu-methodological:2011}. Text-to-speech (TTS) voices are typically designed to synthesise a single most likely rendition of a given sentence. While many methods have been proposed to add control to TTS voices, often they do not take this arbitrary variation into account. In contrast, we focus on designing TTS voices that are able to produce any viable prosodic realisation of a given sentence in isolation. Such a system could be driven by contextual information (e.g.\ provided by a dialogue system) to produce more appropriate prosodic renditions. However, we here focus on the task of producing random (but acceptable) prosodic renditions given an isolated sentence.

Since neural statistical parametric speech synthesis (SPSS) became the leading paradigm in speech synthesis research \cite{blizzard-decade:2014} most TTS voices have used static plus dynamic features optimised with mean squared error, followed by maximum likelihood parameter generation (MLPG) and post-filtering \cite{heiga-DNN-SPSS:2013}. These methods are a legacy of hidden Markov model (HMM) SPSS \cite{heiga-HMM-SPSS:2009}, where the problem of oversmoothing was observed and methods were developed to mitigate it. Oversmoothing of acoustic features is still an issue in neural SPSS, due to a combination of assumptions made in designing models \cite{gustav-REHASP:2014}. Here we focus on prosody (and specifically on modelling intonation, which is the F\textsubscript{0} component of prosody) where the symptom of oversmoothing is flatter, more average prosody.

We argue that a model designed to synthesise distinct renditions will, by design, \emph{not} model average prosody. Variational autoencoders (VAEs) are a class of generative models that can learn a smooth latent space approximating the true latent factors of the data. Therefore, we use a VAE \cite{VAE:2013} to tackle the problem of average prosody, using the latent space to capture otherwise unaccounted-for variation. We propose that by sampling from the low-probability regions of the VAE's prior we can generate idiosyncratic prosodic renditions.

\section{Related work} \label{sec:related_work}

Methods for control of SPSS voices roughly fall into two categories: explicitly labelled control and latent control. The former is typically expensive because labelling is labour-intensive, although this can be automated at the expense of accuracy \cite{AuToBI:2010,zack-IS18:2018}. Labelling requires a concrete and consistent schema that can be followed by human annotators. For many aspects of variation in speech this is challenging, a clear example being emotion labelling \cite{emotion-datasets:2003}. For example, categorical emotions (e.g.\ happy or sad) may be too coarse, and appraisal-based measures (e.g.\ arousal or valence) may be too complex or ambiguous for labellers \cite{zack-MScR:2017}. Additionally, there is the question of elicitation: should natural speech be annotated, or should the variation of interest be elicited (e.g., acted) and assumed to be correct?

It has been shown that unsupervised methods can achieve similar results to supervised control \cite{gustav-GMMQ-VAE:2018}, which may be related to the challenge of accurately labelling variation in real data, as discussed above.

Discriminant condition codes, first proposed for speech recognition \cite{DCC:2014} have proved useful for multi-speaker TTS \cite{input-codes:2017}. The same method has been applied in an unsupervised fashion \cite{oliver-control:2015}, allowing for control of arbitrary variation. While these methods have been shown to have a probabilistic interpretation \cite{gustav-GMMQ-VAE:2018}, they do not model uncertainty or guarantee smoothness in the latent space. As we discuss in Section~\ref{sec:VAE}, this smoothness is important for determining what corresponds to an idiosyncratic (and thus more varied) rendition of a sentence.

Tacotron \cite{tacotron:2017} is a sequence-to-sequence model, for which style control using ``global style tokens'' (GST) has been proposed \cite{tacotron-GST:2018}. GSTs produce high quality speech, and can be predicted from text \cite{tacotron-TP-GST:2018}. However, individual GSTs cannot be effectively used to produce distinct styles as they are trained as weighted combinations; using individual GSTs leads to significantly degraded audio quality. We expect a random weighting of the tokens will also produce degraded naturalness, since there is no smoothness constraint.  % on the GST space.

VAEs have been demonstrated for speech synthesis \cite{hierarchical-VAE:2019,voice-loop-VAE:2019}, voice conversion \cite{hsu2017voice}, and intonation modelling \cite[Chapter~7]{xin-thesis:2018}.
Discrete representations have also been incorporated into the VAE framework \cite{discrete-VAE:2016,VQ-VAE:2017}. An experiment with VQ-VAE \cite{VQ-VAE:2017} demonstrated that phones can be learnt with unsupervised training, a result promising for potentially learning discrete prosodic styles. However, in this work we use a continuous latent space.

The recently-introduced clockwork hierarchical VAE (CHiVE) \cite{CHiVE:2019} is similar to the model we propose here, however our VAE does not make use of the clockwork hierarchical structure and we only predict intonation, while CHiVE predicts F\textsubscript{0}, duration, and C\textsubscript{0}. Since we consider isolated sentences, we are not concerned with a single ``best'' output of our system.

Prior work using VAEs has focused on modelling segmental features \cite{VQ-VAE:2017,hierarchical-VAE:2019}, with some applications to intonation modelling, e.g.\ for style transfer \cite{CHiVE:2019} and predicting latents from text \cite[Chapter~7]{xin-thesis:2018}. 
However, our method moves towards TTS systems that can synthesise multiple distinct prosodic renditions (in an unsupervised framework and without the need for control).

\section{Average prosody} \label{sec:average-prosody}

While many methods have been proposed to add control, there is a more fundamental issue, known as oversmoothing, which leads to flatter, more boring prosody. Typical SPSS uses either feedforward neural networks, or recurrent neural networks (RNNs) to map from a linguistic specification to acoustic features. This mapping is learnt by minimising the mean squared error (MSE) against the ground truth acoustics. MSE is equivalent to minimising the negative log-likelihood (NLL) of a unit-variance Gaussian. This has two effects on such SPSS models: they learn the mean of the data, and are sensitive to outliers. By modelling the mean, SPSS models over-smooth the acoustics -- in the context of prosody this is known as \emph{average prosody}.

Methods such as the $\epsilon$-contaminated Gaussian \cite{heiga-contaminated-normal:2016} exist to handle outliers. However, to fix both issues, it is common to collect speech that is as controlled and consistent as possible in terms of style. Training data with a single style results in models which produce more natural speech \cite{skripto:2018}, but it also limits the voice's stylistic range. If we are interested in producing more varied style/prosody/intonation we need more varied data, but this must then be handled appropriately by our model.

Generative models, such as Mixture density networks (MDN) \cite{MDN:1994}, have the ability to handle multiple modes. MDNs parameterise a Gaussian mixture model (GMM) for each acoustic frame which can help with oversmoothing of spectral features \cite{heiga-MDN:2014}. However, for prosodic features, we are interested in fixing oversmoothing over a longer timescale, for which frame-level GMMs are less suitable. Instead, we use variational autoencoders which model a distribution in an abstract (latent) space at whichever timescale is preferred.

\section{Variational autoencoders} \label{sec:VAE}

Variational autoencoders (VAEs) \cite{VAE:2013} are a class of latent variable models, i.e.\ they learn some unsupervised latent representation of the data. They consist of an encoder and a decoder: the encoder parameterises the approximate posterior $q_{\phib}(\z \mid \x)$, which is an approximation of $p_{\thetab}(\z \mid \x)$ -- the underlying factors that describe the data. The decoder is trained to reconstruct the input signal $\x$ from this latent space, i.e.\ given a sample from the posterior $\tilde{\z} \sim q_{\phib}(\z \mid \x)$, we reconstruct $\bar{\x} \sim p_{\thetab}(\x \mid \tilde{\z})$. The encoder and decoder are trained jointly by maximising the evidence lower bound (ELBO),
\begin{align}
    \mathcal{L}(\thetab, \phib; \x) = -&KL(q_{\phib}(\z \mid \x) \mid\mid p(\z)) \nonumber \\
    &+ \E_{q_{\phib}(\z \mid \x)} \left[ \log p_{\thetab}(\x \mid \z) \right]
\end{align}

The first term in the ELBO enforces a prior on the approximate posterior, while the second term measures reconstruction error. The Kullback-Leibler (KL) divergence term -- used to enforce the prior -- puts a cost on using the latent space. This cost on transmitting information through the latent space can encourage the approximate posterior to collapse to the prior, thus encoding no information: posterior collapse. KL-cost annealing is a common way to mitigate posterior collapse \cite{beta-VAE:2017}, where the KL term is down-weighted at the start of training, reducing the cost of encoding information in the latent space.

Here we consider conditional VAEs \cite{conditional-VAE:2015}, which model F\textsubscript{0} conditioned on linguistic features. We use a sentence-level approximate posterior, although a sequence of phrase- or syllable-level latents would be a reasonable alternative. We use an isotropic Gaussian prior $p(\z) = \mathcal{N}(\z; \mathbf{0}, \mathbf{1})$, which gives an analytical form of the KL term.

Enforcing a Gaussian prior gives another useful quality: the single mode and smooth pdf means the distance of $q_{\phib}(\z \mid \x)$ from the prior mean $\mathbf{0}$ will be inversely proportional to the similarity of $\x$ and the largest mode in the data (e.g., the most common prosodic style). That is, the most idiosyncratic $\x$ will be far from the peak at $\mathbf{0}$. This is helpful for our interest in varied prosodic renditions; we can generate varied prosodic renditions using the decoder by sampling low-density regions in the prior. Thus we define two models that use only the VAE decoder,
\begin{align}
    &\z_{\textsc{peak}} = \mathbf{0} &&\bar{\x}_{\textsc{peak}} \sim p_{\thetab}(\x \mid \z_{\textsc{peak}}) \\
    &\z_{\textsc{tail}} \sim vMF(\kappa=0) &&\bar{\x}_{\textsc{tail}(r)} \sim p_{\thetab}(\x \mid r \times \z_{\textsc{tail}})
\end{align}

\noindent where $\bar{\x}_{\textsc{peak}}$ should correspond to the most common mode, i.e.\ style. Due to the uni-modal prior $p(\x)$, $\bar{\x}_{\textsc{peak}}$ may instead correspond to an average of multiple styles, i.e.\ average prosody. Our proposed model uses $\z_{\textsc{tail}}$ (uniform samples on a hypersphere's surface\footnote{Sampled from a von Mises-Fisher distribution ($vMF$) with uniform concentration -- \href{https://en.wikipedia.org/wiki/Von\_Mises–Fisher\_distribution}{wikipedia.org/wiki/Von\_Mises–Fisher\_distribution}}) to produce idiosyncratic renditions $\bar{\x}_{\textsc{tail}(r)}$, where the larger the radius $r$ is the more unlikely the rendition.

\section{Systems}

We focus on modelling intonation, though in the future we plan to extend this to complete prosodic modelling (F\textsubscript{0}, duration and energy). Modelling only F\textsubscript{0} limits the range of variation we can achieve, but reduces the risk of producing unnatural speech: spectral features and durations are taken from natural speech in our experiments, with full TTS left for future work. We use the WORLD vocoder \cite{WORLD:2016}, for analysis and synthesis. Our models\footnote{Code is available at \href{https://github.com/ZackHodari/average_prosody}{github.com/ZackHodari/average\_prosody}} are implemented in PyTorch \cite{pytorch:2017}.

\begin{figure}[t]
  \centering
  \includegraphics[width=\linewidth]{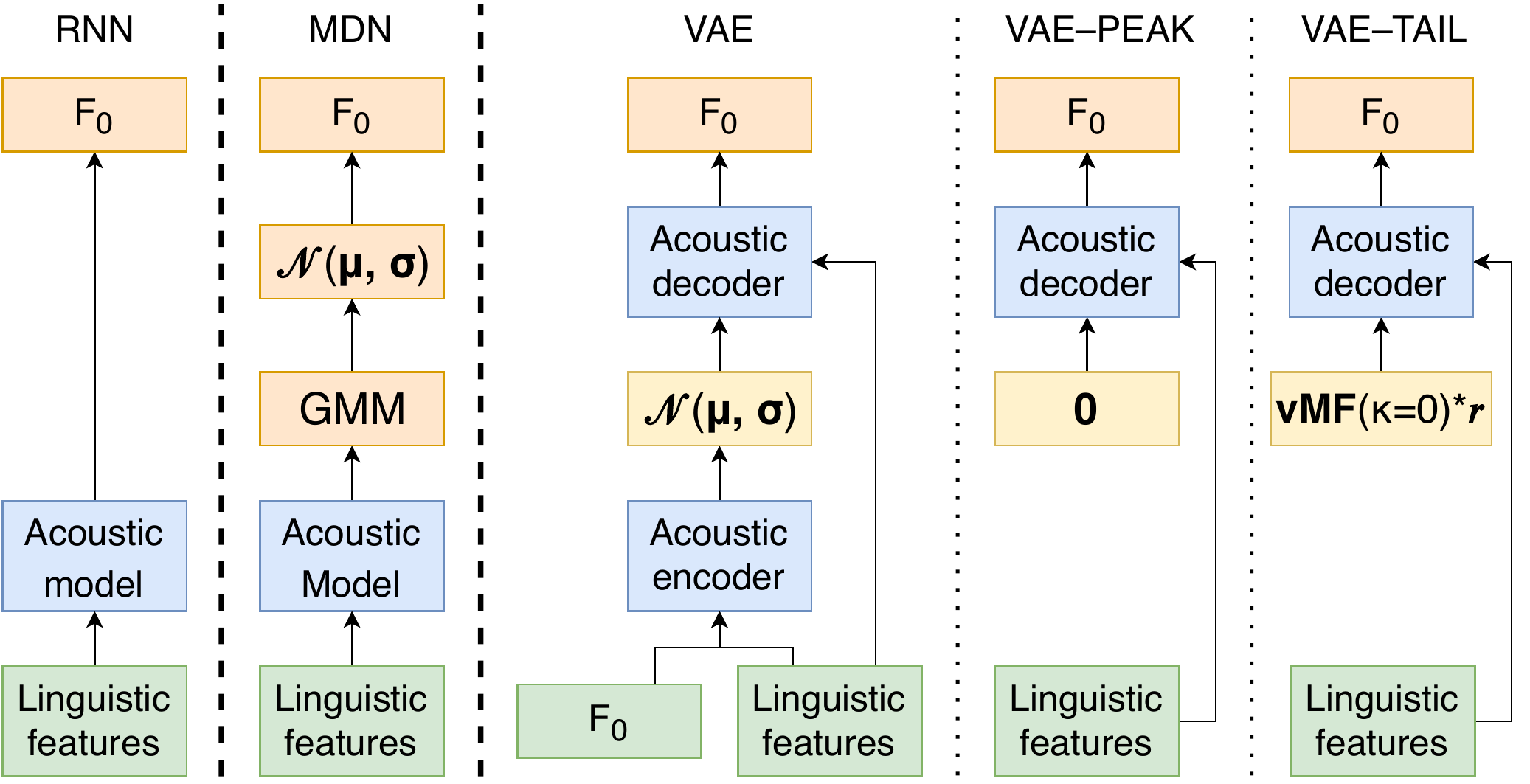}
  \caption{Illustration of our models, where only the first three are trained models. \textsc{vae--peak} and \textsc{vae--tail} are different configurations of the VAE model. Blue: learned modules. Green: frame-level inputs. Orange: frame-level predictions. Yellow: sentence-level latent space.}
  \label{fig:models}
  \vspace{-12pt}
\end{figure}

We use the same basic recurrent architecture for all trainable modules in Figure~\ref{fig:models}: a feedforward layer with 256 units, followed by three uni-directional recurrent layers using gated recurrent cells (GRUs) \cite{GRU:2014} with 64 units, finally any outputs used are projected to the required output dimension.

We use 600-dimensional linguistic labels from the standard Unilex question-set and 9 frame-level positional features with min-max normalisation as in the standard Merlin recipe \cite{merlin:2016}. The model predicts logF\textsubscript{0}, delta (velocity), and delta-delta (acceleration) features with mean-variance normalisation. We use Adam \cite{adam:2014} with an initial learning of 0.005, which is increased linearly for the first 1000 batches, and then decayed proportional to the inverse square of the number of batches \cite[Sec~5.3]{transformer:2017}, where our batch size is 32. Early stopping is used based on validation performance. MLPG \cite{MLPG:2000} is used to generate the F\textsubscript{0} contour from the dynamic features; predicted standard deviations are used by the MDN, and all other models use the global standard deviation of the training data.

The MDN uses four mixture components, whose variances are floored at $10^{-4}$. To synthesise from the MDN, we use the most likely component sequence (i.e.\ argmax) to select means and variances used in MLPG.

Systems \textsc{vae--peak} and \textsc{vae--tail} in the list below are identical apart from the use of different sampling schemes (see Section~\ref{sec:VAE}). Their shared model uses a 16-dimensional isotropic Gaussian as the approximate posterior. The latent sample $\tilde{\z}$ is broadcast to frame-level and input to the decoder, along with the linguistic features. The decoder predicts static and dynamic logF\textsubscript{0} features; as such the reconstruction loss is MSE. The KL-divergence term is weighted by zero during the first epoch and increased linearly to 0.01 over 40 epochs. Using this annealing schedule, the model converged to a KL-divergence of 3.13.

\begin{enumerate}[leftmargin=2cm,itemsep=2pt,topsep=5pt,parsep=0pt,partopsep=0pt]
    \item[\textsc{rnn}        ] Standard RNN-based SPSS model, using MSE.
    \item[\textsc{mdn}        ] MDN with 4 mixture components, using NLL.
    \item[\textsc{vae--peak}  ] VAE decoder using $\z_{\textsc{peak}}$, i.e.\ the zero vector.
    \item[\textsc{vae--tail}  ] VAE decoder using $\z_{\textsc{tail}}$ with $r=3$, i.e.\ points on the surface of a hypersphere with radius 3.

    \item[\textsc{copy--synth}] Natural F\textsc{0}.
    \item[\textsc{baseline}   ] A quadratic polynomial fitted to natural F\textsubscript{0}.
    \item[\textsc{rnn--scaled}] F\textsubscript{0} from \textsc{rnn}, scaled vertically by a factor of 3.
\end{enumerate}

\subsection{Purpose of baselines}

\textsc{baseline} sets a lower bound on naturalness (and variedness): no matter how much variation a system produces, its naturalness should never fall below that of \textsc{baseline}.
An upper-bound is \textsc{copy--synth}: no system should be more natural than this, but might sound more varied even though it is unclear whether this would be favoured by listeners.

Because we expect that adding more variation will degrade naturalness, we wish to quantify this. \textsc{rnn--scaled} is intended as a lower-bound on naturalness using a na\"ive method for increasing variation, similar to variance scaling \cite{hanna-GV:2012}.
\textsc{rnn--scaled} is intended to demonstrate that \textsc{vae--tail} can produce the same amount of perceived variation but \emph{without sacrificing} as much naturalness.
In this study, setting the amount of perceived variation in \textsc{rnn--scaled} and \textsc{vae--tail} was calibrated in a pilot listening test by the authors, where we attempted to match the level of variation to \textsc{copy--synth}.

\section{Hypotheses} \label{sec:hypotheses}

\begin{enumerate}[leftmargin=0.6cm,itemsep=2pt,topsep=5pt,parsep=0pt,partopsep=0pt]

    \item[\textbf{H\textsubscript{1}}] \textsc{vae--tail} will be much more \textbf{varied} than the typical SPSS systems (\textsc{rnn}, \textsc{vae--peak}, \textsc{mdn}).
    
    \item[\textbf{H\textsubscript{2}}] \textsc{rnn--scaled}, \textsc{vae--tail}, and \textsc{copy--synth} will have the same level of \textbf{variedness}.
    
    \item[\textbf{H\textsubscript{3}}] \textsc{rnn}, \textsc{vae--peak}, and \textsc{mdn} will have a similar level of \textbf{variedness}, where \textsc{mdn} is more varied than the other two.

    \item[\textbf{H\textsubscript{4}}] \textsc{vae--tail} will have slightly lower \textbf{naturalness} than the typical SPSS systems (\textsc{rnn}, \textsc{vae--peak}, \textsc{mdn}).
    
    \item[\textbf{H\textsubscript{5}}] \textsc{vae--tail} will be much more \textbf{natural} than the varied baseline \textsc{rnn--scaled}.

\end{enumerate}

\section{Data} \label{sec:data}

Our choice of training data is motivated by the need for prosodic variation: if the data is very stylistically consistent, there will be too little variation for the VAE to capture in its latent space. We therefore use the Blizzard Challenge 2018 dataset \cite{blizzard:2018} provided by Usborne Publishing. The data consists of stories read in an expressive style for a 4--6 year old audience, with some character voices. Many of the stories include substantial amounts of direct speech. In total it contains 6.5 hours (\texttildelow7,250 sentences) of professionally-recorded speech from a female speaker of standard southern British English. The training-validation-test split described in Watts et al.\ \cite{oliver-control:2015} is used.

\section{Evaluation} \label{sec:evaluation}

We want to evaluate the amount of variation produced by the systems described. However, variation alone is not a guarantee of ``better'' speech synthesis \cite{javier-prosody:2014}. For this reason we evaluate quality along with variation. To determine quality we measure naturalness using a standard mean opinion score test, where users were asked to ``rate the naturalness'' on a 5-point Likert scale.

Evaluating variation is less straightforward. We employed a preference test where two systems were compared side by side for the same sentence. Users were asked to choose ``which sentence has more varied intonation'', where one sentence must be be marked as ``more flat'', and the other as ``more varied''. %, this is a forced choice. 
Due to the large number of pairs for 7 systems, we excluded \textsc{baseline} in the pairwise test, as it is clear from the speech samples\footnote{Speech samples available at \href{https://github.com/ZackHodari/average_prosody}{github.com/ZackHodari/average\_prosody}} that it would be the least varied. However, without \textsc{baseline} in the variation test we lose our lower-bound on variation.

We randomly selected 32 test sentences of between 7 and 11 words (1.4 to 4.8 seconds). The naturalness test was performed before the preference test. As there were 22 screens to be completed for each sentence it was necessary to split the test into two halves using a simple 2x2 Latin square between-subjects design. In total we used 30 participants, 15 per listener group, the test took 45 minutes and participants were paid \pounds8.

\section{Results} \label{sec:results}

\subsection{Naturalness test} \label{sec:results-MOS}

\begin{figure}[t]
  \centering
  \includegraphics[width=\linewidth]{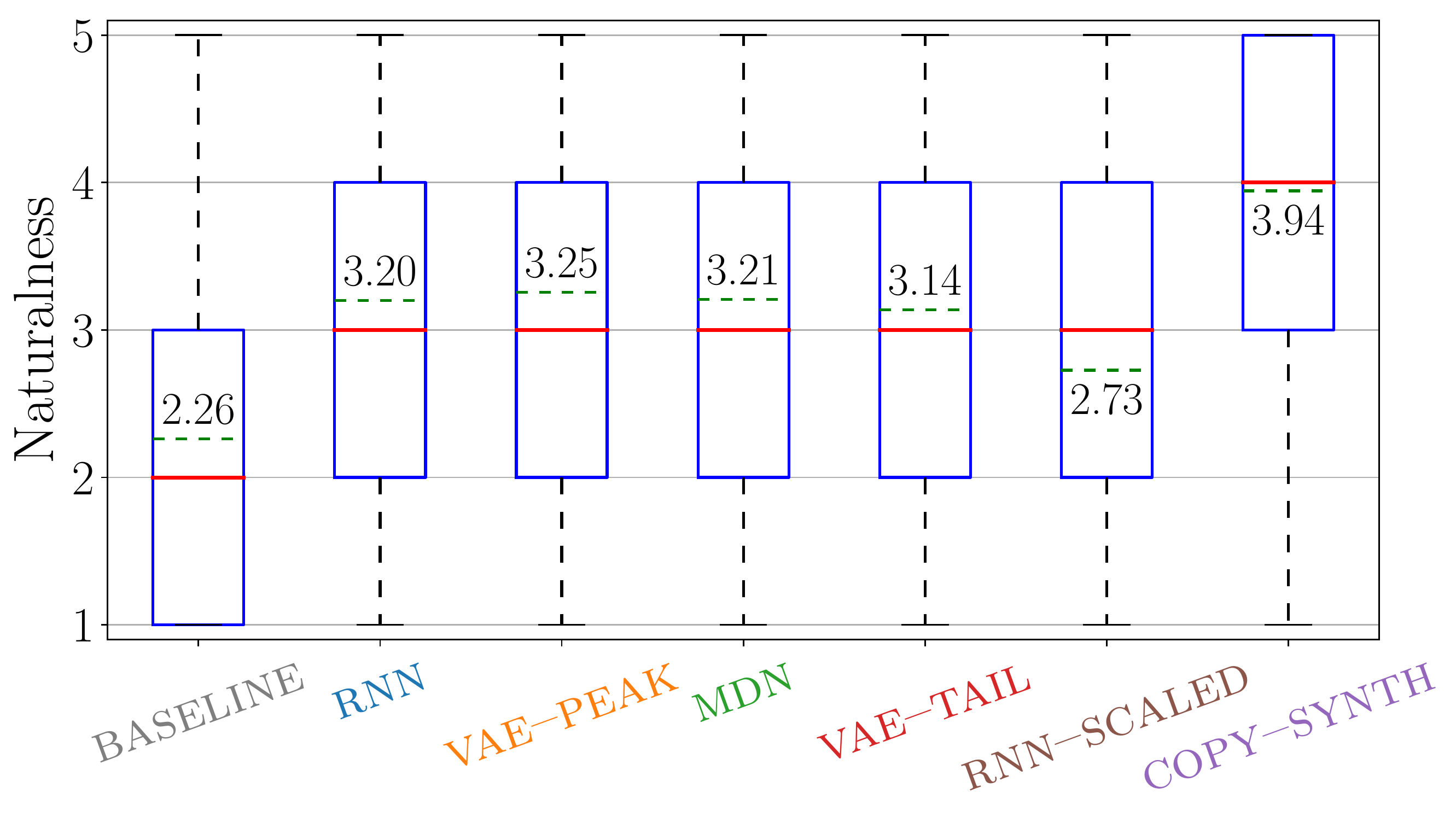}
  \caption{Naturalness results. Solid red lines are medians, dashed green lines are means (cannot be used for statistical comparison),  blue boxes show the 25\textsuperscript{th} and 75\textsuperscript{th} percentiles, and whiskers show the range of the ratings, excluding outliers which are plotted with $+$. Ordered according to the variation test.}
  \label{fig:MOS}
\end{figure}

A summary of the naturalness ratings is provided in Figure~\ref{fig:MOS}. We perform a Wilcoxon rank-sums significance test between all pairs of systems in the naturalness test, followed by Holm-Bonferroni correction. This statistical analysis is the same as for the Blizzard challenge \cite{blizzard-analysis:2007}. \textsc{vae--tail}, \textsc{rnn}, \textsc{mdn}, and \textsc{vae--peak} form a group for which we did not find any significant difference. All other system pairs are significantly different, with a corrected p-value of less than 0.00001.

\subsection{Variation test} \label{sec:results-variation}

\begin{figure}[t]
  \centering
  \includegraphics[width=\linewidth]{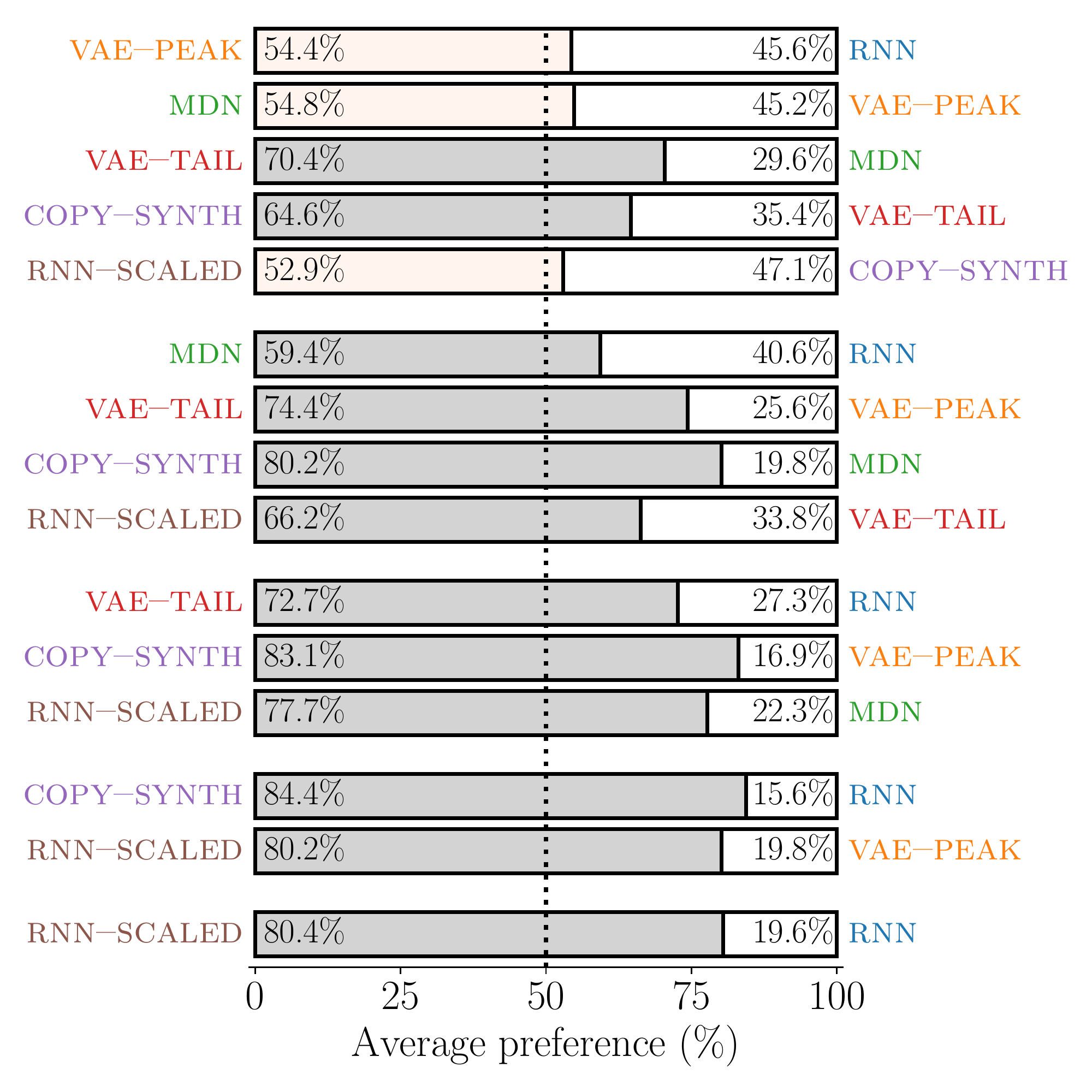}
  \caption{Pairwise variedness results. Pairs are ordered such that the more varied system is on the left. The top 5 rows give the pairs that are consecutive in the ordering, with following rows showing systems that are increasingly further apart in the ordering. We did not find a significant difference for the pairs marked in a lighter colour.}
  \label{fig:AB}
\end{figure}

While it is not guaranteed that human preferences are self-consistent, or globally consistent\footnote{As described by Arrow's impossibility theorem \cite{arrows-impossibility:1950}.}, we see that the results in Figure~\ref{fig:AB} do form a consistent ordering from most flat to most varied: \textsc{rnn} $\rightarrow$ \textsc{vae--peak} $\rightarrow$ \textsc{mdn} $\rightarrow$ \textsc{vae--tail} $\rightarrow$ \textsc{copy--synth} $\rightarrow$ \textsc{rnn--scaled}. However, relative variedness is sometimes inconsistent, e.g.\ while \textsc{rnn--scaled} is more varied than \textsc{copy--synth} (5\textsuperscript{th} row), we see that the difference between \textsc{copy--synth} and \textsc{rnn} (13\textsuperscript{th} row) is greater than the difference between \textsc{rnn--scaled} and \textsc{rnn} (15\textsuperscript{th} row).

We perform a binomial significance test for the 15 pairs in the listening test, followed by Holm-Bonferroni correction.
With the correction we find that (\textsc{rnn}, \textsc{vae--peak}), (\textsc{vae--peak}, \textsc{mdn}), and (\textsc{copy--synth}, \textsc{rnn--scaled}) did not show a significant difference: this is indicated by the colouring of those pairs in Figure~\ref{fig:AB}. All other pairs are significantly different, with a corrected p-value of less than 0.0002.

\subsection{Naturalness--Variedness trade-off} \label{sec:results-relative}

While this ordering supports our expectations, we cannot clearly comment on their support of our hypotheses in Section~\ref{sec:hypotheses} as the relative variedness between systems is not clear. Additionally, we would like to clearly compare the trade-off between increasing intonation variation and naturalness. This requires us to represent the pairwise preferences in Figure~\ref{fig:AB} along a single axis.

\begin{figure}[t]
  \centering
  \vspace{-6pt}
  \includegraphics[width=\linewidth]{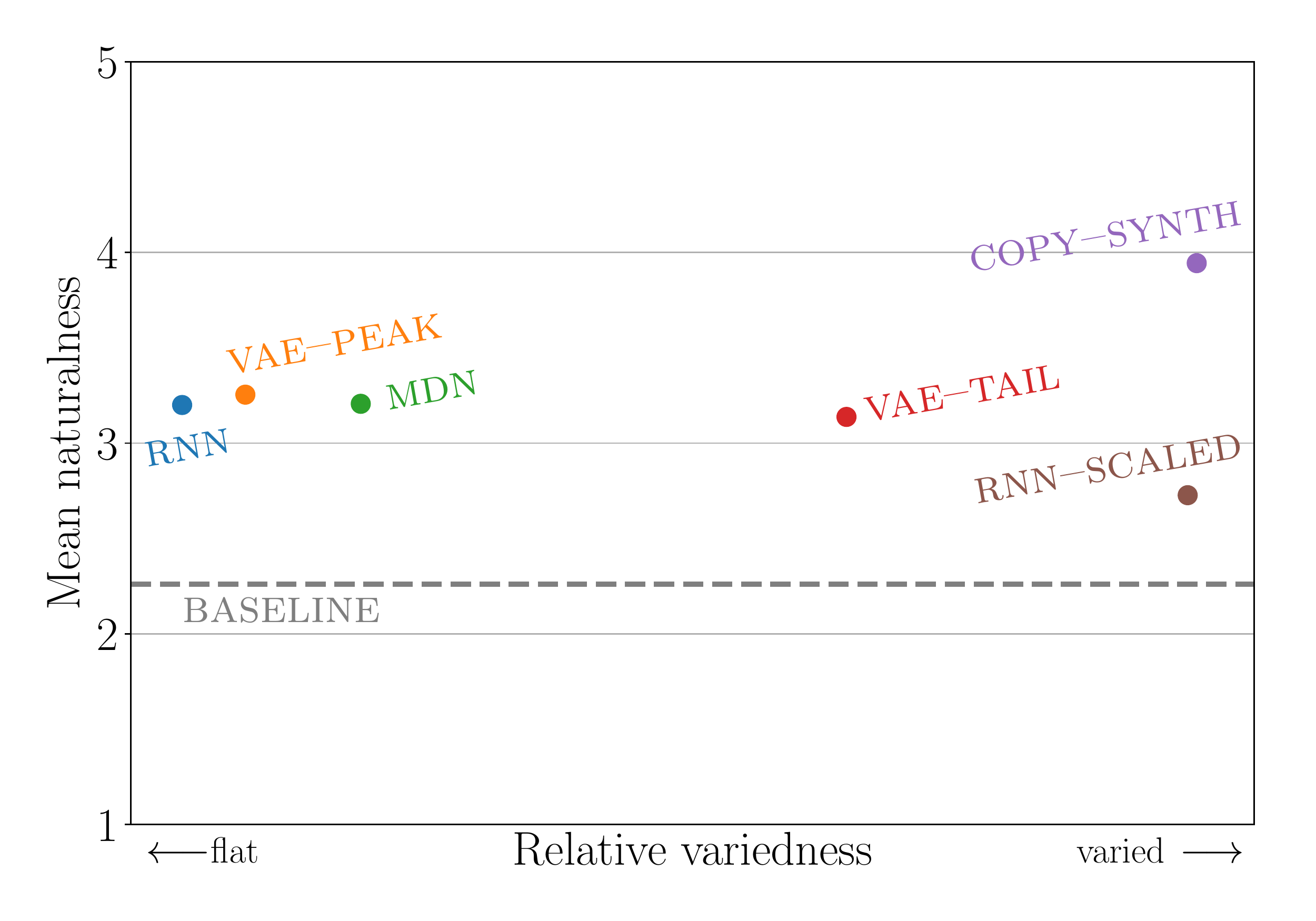}
  \caption{Naturalness-variedness trade-off. Ideally as we increase the amount of prosodic variation our system will not decrease in naturalness. Note that naturalness comparisons can only be made using the significance results in Section~\ref{sec:results-MOS}.}
  \label{fig:MOS-vs-variation}
\end{figure}

We could approach this using multi-dimensional scaling (MDS) \cite{MDS:2003}; however, the pairwise preferences correspond to directed edges, not distances. Instead, we formulate the problem as a system of linear equations\footnote{We thank Erfan Loweimi and Gustav Henter for insightful discussions that led to this formulation of the problem.}. Here, the variables are the positions of each system in the dimension of relative variedness, and each equation describes the ``excess preference'' of a system pair (the difference between the two system's average preference). This system can be solved using ordinary least squares:
$$Ax = b \qquad\qquad x = (A^TA)^{-1}A^Tb$$

\noindent where $A \in \{-1, 0, 1\}^{15 \times 6}$ and $b \in \mathbb{R}^{15 \times 1}$ encode the pairwise results in Figure~\ref{fig:AB}. Given the solution ($x \in \mathbb{R}^{6 \times 1}$) we plot naturalness against relative variedness in Figure~\ref{fig:MOS-vs-variation}. Systems to the left have flatter intonation, and systems to the right have more varied intonation. This axis represents human preference and is not intended to be a perceptual scale.

In Figure~\ref{fig:MOS-vs-variation}, we see that \textsc{vae--tail} is much more varied than the typical SPSS systems (\textbf{H\textsubscript{1}}). It is also clear that our calibration favoured less variation in \textsc{vae--tail} than \textsc{copy--synth} (rejecting \textbf{H\textsubscript{2}}), thus we cannot make broad statements about the naturalness-variedness trade-off. However, based on the significant drop in naturalness from \textsc{rnn} to \textsc{rnn--scaled}, and the clustering over relative variedness, we believe that \textsc{vae--tail} would still be significantly more natural than \textsc{rnn--scaled} even if it matched \textsc{copy--synth}'s level of variation.

\textsc{rnn}, \textsc{vae--peak}, and \textsc{mdn} are clustered along the axis of relative variation, with \textsc{mdn} being significantly more varied, but only by a small amount (\textbf{H\textsubscript{3}}). Demonstrating that all systems suffer from oversmoothing of F\textsubscript{0} to a similar extent.

While the mean naturalness of \textsc{vae--tail} is lower than \textsc{rnn}, \textsc{vae--peak}, and \textsc{mdn}, the means cannot be directly compared, and no significant difference was found in Section~\ref{sec:results-MOS}. Rejecting \textbf{H\textsubscript{4}} suggests we can produce more varied intonation without sacrificing naturalness. However, we expect that with the ideal calibration we may see some slight degradation in naturalness of \textsc{vae--tail}. We do observe that \textsc{vae--tail} is much more natural than \textsc{rnn--scaled} (\textbf{H\textsubscript{5}}).

\begin{figure*}[t]
  \centering
  \includegraphics[width=\linewidth]{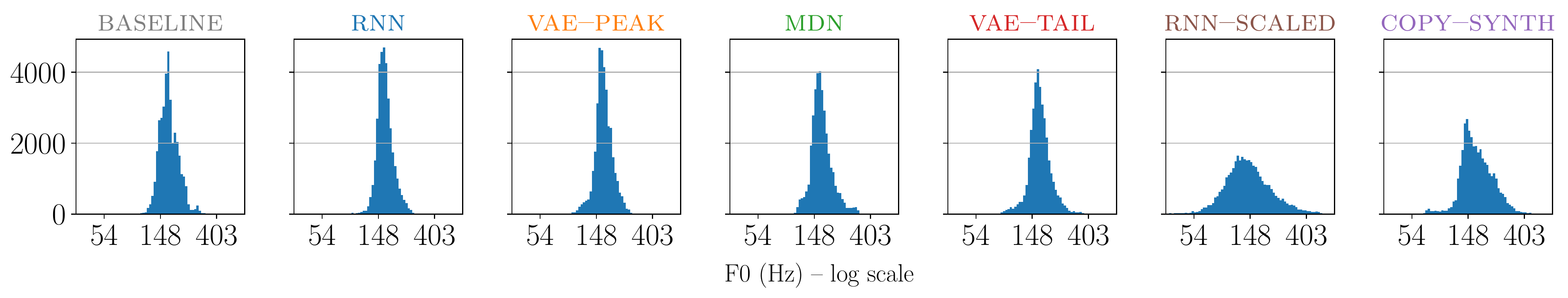}
  \caption{Histogram of logF\textsubscript{0} values for each system over all the listening test material. Ordered according to the variation test.}
  \label{fig:F0-hist}
  \vspace{-1pt}
\end{figure*}

\section{Analysis} \label{sec:analysis}

\paragraph*{Calibration} The horizontal axis in Figure~\ref{fig:MOS-vs-variation} shows \textsc{vae--tail} having much greater perceived intonation variation than \textsc{mdn}, while the logF\textsubscript{0} histograms in Figure~\ref{fig:F0-hist} shows them as having the same amount of objective variation -- variance of logF\textsubscript{0} predictions for the listening test stimuli. This demonstrates that objective measures do not necessarily correspond to perceived variation, which is exactly what makes calibration of \textsc{vae--tail} and \textsc{rnn--scaled} difficult. Figure~\ref{fig:F0-hist} shows that \textsc{vae--tail} has a narrower histogram than \textsc{copy--synth}, however as objective measures do not necessarily correspond to perceived variation we chose not to rely on objective measures for calibration.

\begin{figure}[t]
  \centering
  \includegraphics[width=\linewidth]{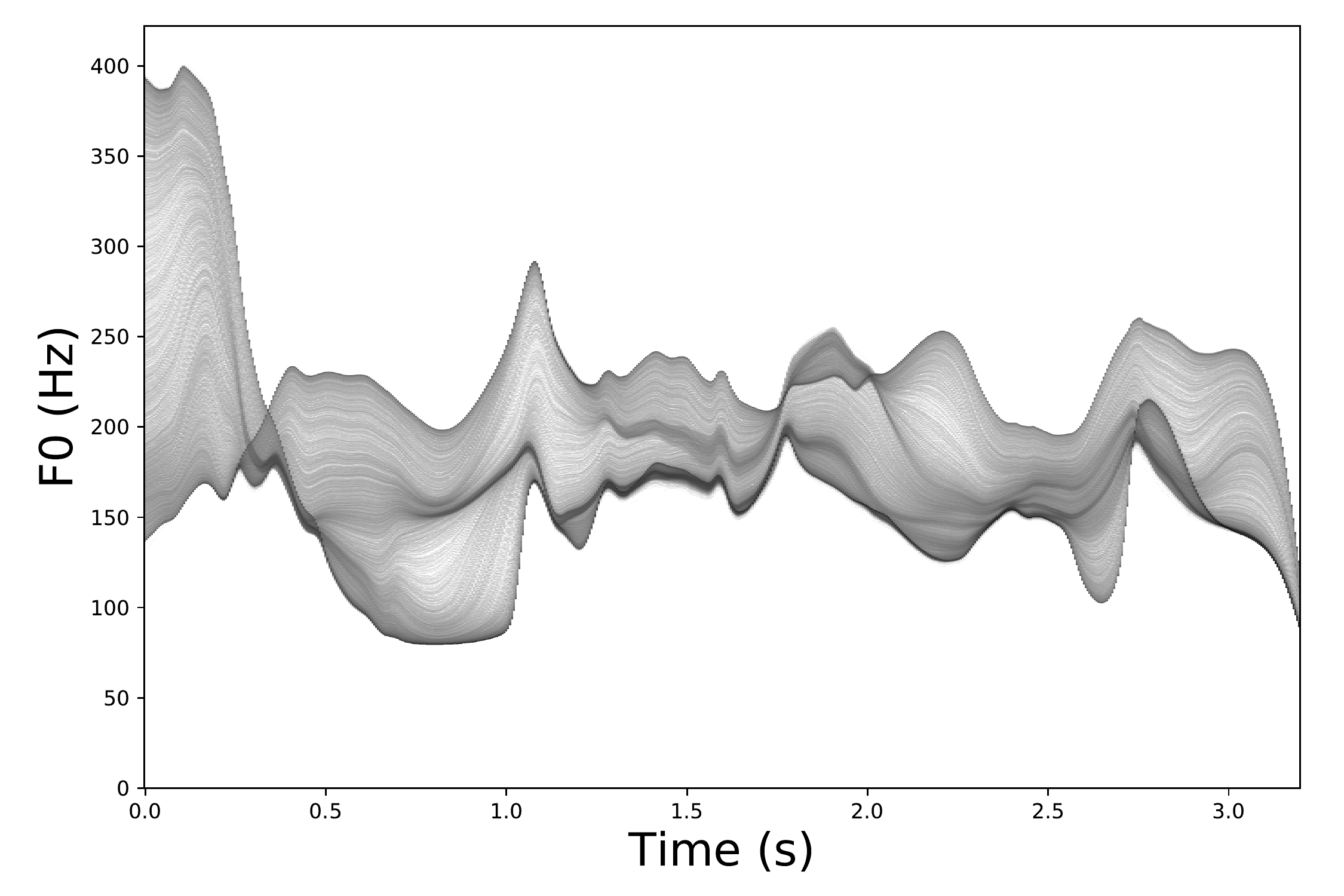}
  \caption{Density of F\textsubscript{0} predictions made by \textsc{vae--tail} for the sentence "Goldilocks skipped around a corner and saw..."}
  \label{fig:F0-density}
\end{figure}

\paragraph*{Multiple renditions} We have demonstrated the ability to produce varied intonation while maintaining the same level of naturalness, thus mitigating average prosody. However, we have not demonstrated \textsc{vae--tail}'s ability to produce multiple distinct prosodic renditions. In Figure~\ref{fig:F0-density} we present a density of 10,000 F\textsubscript{0} contours $\bar{\x}_{\textsc{tail}(3)}$ produced using samples $\z_{\textsc{tail}} \sim vMF(\kappa=0)$. As expected, the F\textsubscript{0} contours produced vary smoothly, but more importantly we see that they vary between multiple distinct contours. For this sentence we see that there may be three distinct contours. We are interested in evaluating the distinctiveness of multiple different samples from \textsc{vae--tail}; however, this is out of the current scope.

\paragraph*{MDN sampling} While \textsc{mdn} is also a generative model, sampling from the frame-level GMMs is not straightforward. MLPG can be used to select the single best trajectory \cite[Case~3]{MLPG:2000}. But to produce multiple renditions from \textsc{mdn} we must choose a sequence of Gaussian components. However, randomly choosing components produces noisy F\textsubscript{0} contours, and using the same component for the entire sequence does not produce distinct performances. This is likely because the components don't represent modes of the data, but behave in a similar way to the $\epsilon$-contaminated Gaussian distribution \cite{heiga-contaminated-normal:2016}.

\section{Conclusion} \label{sec:conclusion}

We have demonstrated that output from typical RNN and MDN models exhibits flat intonation. Additionally, we have provided evidence that sampling from the tails of a VAE prior produces speech that is much more varied than typical SPSS while maintaining the same level of naturalness. In future we plan to undertake a full evaluation of this trade-off, to determine if and when this method begins to improve or degrade in quality.

In future work, we plan to: use MUSHRA in place of a preference test; use a neural vocoder; make use of seq2seq models with attention instead of upsampling the linguistic features; predict other prosodic features;
and make use of either a discrete latent space \cite{discrete-VAE:2016} or a mixture model VAE prior \cite{VampPrior-VAE:2018}.

{\footnotesize \noindent \textbf{Acknowledgements:} Zack Hodari was supported by the EPSRC Centre for Doctoral Training in Data Science, funded by the UK Engineering and Physical Sciences Research Council (grant EP/L016427/1) and the University of Edinburgh. Oliver Watts was supported by EPSRC Standard Research Grant EP/P011586/1.}

% \singlespace
\bibliographystyle{IEEEtran}
\bibliography{paper_arxiv}

\end{document}